# Molecular rectifiers with very high rectification ratio enabled by oxidative damage in double-stranded DNA

Abhishek Aggarwal[1], Supriyo Naskar[1], and Prabal K. Maiti[1, *]

[1]Center for Condensed Matter Theory, Department of Physics, Indian Institute of Science, Bangalore 560012

*Correspondence— maiti@iisc.ac.in

**Abstract—** In this work, we report a novel strategy to construct molecular diodes with a record tunable rectification ratio of as high as $10^6$ using oxidatively damaged DNA molecules. Being exposed to several endogenous and exogenous events, DNA suffers constant oxidative damages leading to oxidation of guanine to 8-Oxoguanine (8oxoG). Here, we study the charge migration properties of native and oxidatively damaged DNA using a multiscale multiconfigurational methodology comprising of molecular dynamics, density functional theory and kinetic Monte Carlo simulations. We perform a comprehensive study to understand the effect of different concentrations and locations of 8oxoG in a dsDNA sequence on its CT properties and find tunable rectifier properties having potential applications in molecular electronics such as molecular switches and molecular rectifiers. We also discover the negative differential resistance properties of fully oxidized Drew-Dickerson sequence. The presence of 8oxoG guanine leads to the trapping of charge, thus operates as a charge sink, which reveals how oxidized guanine saves the rest of the genome from further oxidative damage.

# Section 1. Introduction

Recent advances in the field of molecular electronics have revealed several fascinating phenomena of quantum-mechanical tunneling regime, such as negative differential resistance[1,2], molecular rectification[3,4], thermoelectric effects[5], and quantum interference[5,6]. The idea of molecular rectification was first anticipated by Aviram and Ratner in 1974[7]. Since then, a plethora of research works have been done to improve the rectification ratio (ratio of current in forward bias and reverse bias) of molecular devices[3,5,8–12]. The rectifying properties of molecules are usually induced using asymmetric molecule-electrode couplings[13], using intercalations in molecular systems[8], or preparing asymmetric molecules[3]. A molecular diode with high rectification ratio (RR) greater than 10 is difficult to form, and only a few works[3,4,14] have reported $RR > 10^2 - 10^3$. In this work, we demonstrate a strategy to obtain molecular rectifiers with a record RR of as high as 6 orders of magnitude using oxidatively-damaged double-stranded (ds) DNA molecules.

In the quest of the miniaturization of electronic devices, dsDNA has emerged as a promising candidate to replace the conventional materials[15–17]. Charge migration in dsDNA is a vast research area leading to the development of several practical applications, from the construction of nanoscale biosensors to an enzymatic tool to detect damage in the genome[15,17–28]. dsDNA molecules inevitably suffer oxidative damages by the reactive oxidative species (ROS)[29,30] causing several DNA lesions and diseases such as cancer and holds a huge biological significance[30–34]. The conversion of guanine to 8-oxoguanine (8oxoG) is one common defect produced by the oxidative damage of dsDNA[29,35]. An increasing body of works suggested that the presence of 8-oxoG in DNA sequences acts as a hole sink, which biologically prevents the rest of the genome from further damage[36,37]. Only a few studies exist in the literature which aim to understand the molecular and electronic structures of oxidatively damaged dsDNA[38–43], but none of them provides an account for the charge dynamics in oxidized dsDNA. This makes the study of the charge transport (CT) in oxidatively-damaged dsDNA immensely important.

Through this work, our aim is to enhance the fundamental understanding behind the CT mechanism in oxidized dsDNA, which helps to understand the physics behind several interesting physical phenomena prevalent in nucleic acid systems, namely rectification and negative differential resistance. Here, we reveal the rectifying properties of oxidatively-damaged dsDNA molecules with a giant RR of as high as $10^6$. Thus, we present a novel strategy to fabricate molecular rectifiers using oxidatively-damaged dsDNA molecules, pushing forward the ever-developing field of molecular electronics.

Using a multiscale multiconfigurational methodology involving atomistic MD simulations, quantum mechanical DFT calculations, and Kinetic Monte Carlo (KMC) simulations, we study the hole migration phenomena in different 12 base pairs (bp) native and 8oxoG-containing dsDNA sequences. By varying the concentration and location of 8oxoG in Drew-Dickerson sequence, we report a rectification ratio of as high as 6 orders of magnitude. In the next section, we describe the multiscale multiconfigurational methodology employed to obtain the CT properties of various nucleic acid systems in this study. This methodology has been employed in several works[17,44–48] and aided in revealing several important aspects of CT in DNA systems.

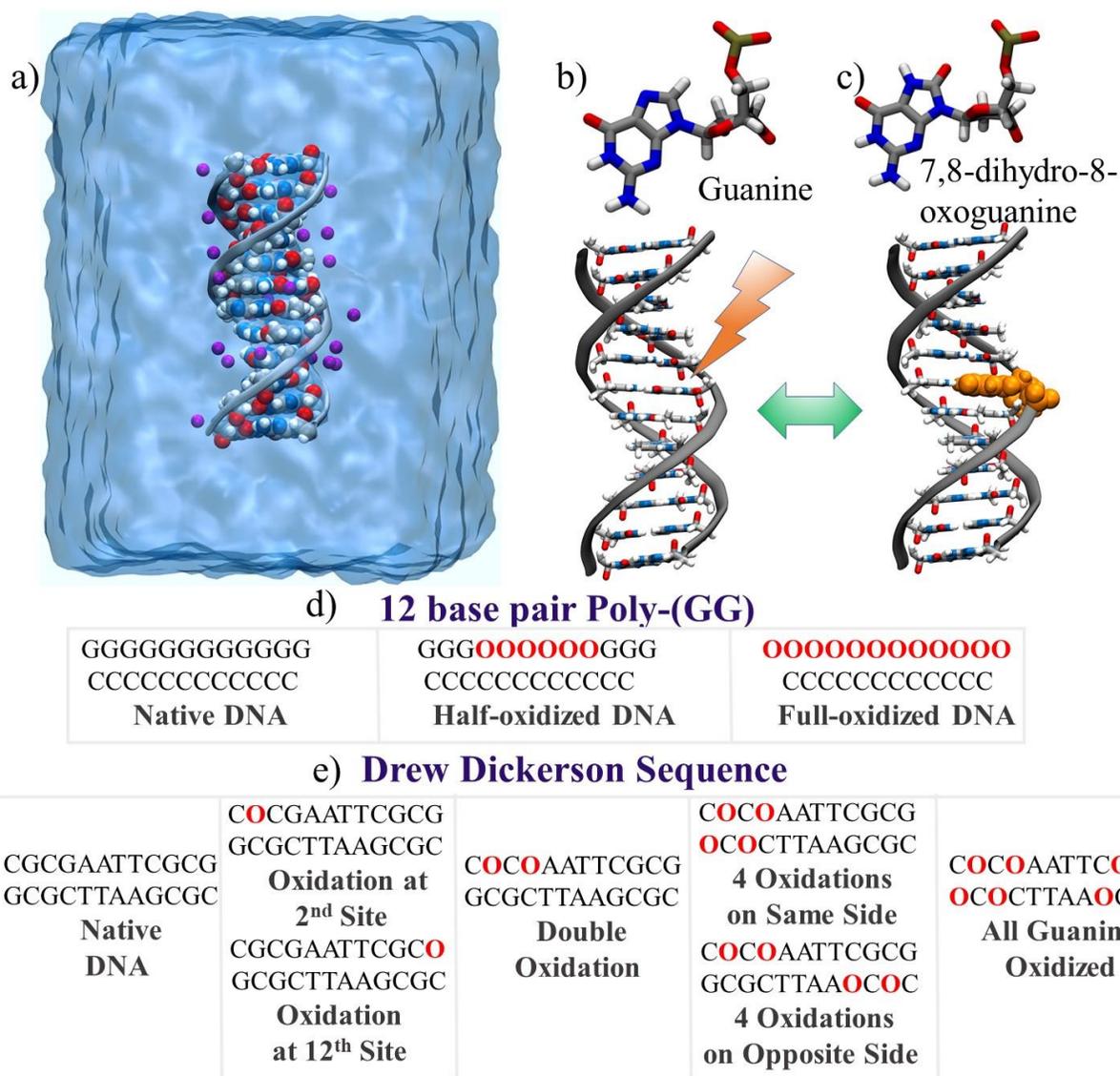

**Fig. 1** Details of the systems studied in this work. (a) Atomistic model of 12 bp oxidatively damaged dsDNA, solvated in water during the MD simulations. For better clarity, the water is shown as a continuous medium, rather than atomistic representation. (b) Atomic representation of guanine and 12 bp native dsDNA, which due to external factors oxidize to (c) 7,8-dihyro-8-oxoguanine (8oxoG) leading to an oxidatively damaged dsDNA. (d) – (e) List of dsDNA sequences simulated in this study. The bold red O represents the replacement of guanine with 8oxoG.

## Section 2. Methodology

### 2.1 MD Simulation Details

To understand the variations of CT properties in a dsDNA sequence when a guanine is replaced by 8oxoG, we first simulate various biologically relevant dsDNA sequences for 100 ns using all-atomistic MD simulations using AMBER18 software package[49], as shown in **Fig. 1 (a)-(c)**. We study 12 bp long poly-G sequence as an extreme case where all and half of the guanine gets oxidized **Fig. 1 (d)**. To study the effect of different concentration and location of 8oxoG, we simulate Drew-Dickerson sequence, oxidized in various manners, as shown in **Fig. 1 (e)**. Such controlled oxidation of guanines in dsDNA has been done in various *in vitro* studies to study their various biological aspects[50,51]. We build bare dsDNA in the B-form for 12 bp poly-G (d-(GGGGGGGGGGGG)) and Drew-Dickerson sequences ((d-(CGCGAATTCGCG))) using the Nucleic Acid Builder (NAB)[52] tool. The oxidized guanine molecule is prepared using the parameters provided in work by Miller et al.[53] We then use the xleap module of the AMBER17[52] tools to solvate each complex in a large rectangular box with TIP3P[54] water model. Charge neutrality of the simulation box is maintained by adding appropriate numbers of Na+ and Cl– ions, for which Joung/Cheatham ion parameters[55] are used.

The system is then energy minimized using the steepest descent method for 1000 steps followed by 2000 steps of the conjugate gradient method keeping a strong harmonic restraint of force constant 500 kcal mol$^{-1}$ Å$^{-1}$ on dsDNA to allow the water molecules to equilibrate. Then the dsDNA is slowly released into the water by reducing the harmonic restraint on it from 20 kcal mol$^{-1}$ Å$^{-1}$ to 0 kcal mol$^{-1}$ Å$^{-1}$ in 5 cycles of steepest descent minimization of 1000 steps each. After energy minimization, the system is slowly heated to reach a temperature of 300 K in 40 ps at a constant pressure of 1 bar. To control the temperature, we use the Langevin thermostat, with collision frequency constant of 2.0 ps$^{-1}$. The Particle Mesh Ewald method[56] (PME) is used to calculate electrostatic interactions. NVT production run is then performed at temperature 300 K and pressure 1 bar for 100 ns. We use the AMBER ff99bsc0 and OL15 force fields for modelling dsDNA molecules. The force-field parameters of 8oxoG are adapted from the study by Miller et al.[57] Bonds involving hydrogen atoms are constrained using the SHAKE algorithm which allowed the use of 2 fs time step. The PMEMD module of the AMBER18[52] software is used for performing the MD simulations.

## 2.2 Hopping Transport Calculations

In this work, the semi-classical Marcus–Hush formalism[45,58] is used to obtain the I-V characteristics of the dsDNA system. In this theory, charge transport is described as incoherent hopping of charge carriers between charge hopping sites. Several previous theoretical and experimental studies have shown that charge transport in dsDNA systems is mediated through strong π-π stacking of nucleobases[17,19,45,59]. Thus, we use DNA nucleobases as hopping sites in this work by replacing the DNA backbone atoms with hydrogen atom for further calculations and optimizations. We use nearest neighbor hopping model where a charge present at any nucleobase other than the terminal bases has 5 available sites to hop to consisting of 4 hopping sites from two adjacent base pairs and the complementary base of the present site.

In Marcus-Hush formalism, the charge hopping rate $\omega_{ik}$ from charge hopping site, $i$, to the hopping site, $k$, is given by[45,58]

$$\omega_{ik} = \frac{2\pi |J_{ik}|^2}{h} \sqrt{\frac{\pi}{\lambda k_B T}} \exp\left[-\frac{(\Delta G_{ik} - \lambda)^2}{4\lambda k_B T}\right] \quad (1)$$

Where $J_{ik}$ is the electronic coupling, also called transfer integral, defined as,

$$J_{ik} = <\Psi^i | H_{ik} | \Psi^k >$$

(2)

Here, $\Psi^i$ and $\Psi^k$ are diabatic wave functions localized on the sites $i$ and $k$, respectively. $H_{ik}$ is the Hamiltonian for the two-site system between which the charge transfer takes place. λ is the reorganization energy. $\Delta G_{ik}$ is the free energy difference between the two sites, $h$ is the Plank's constant, $k_B$ is the Boltzmann constant, and $T$ is the absolute temperature. The electronic couplings between all possible nearest neighbor charge hopping base pairs are computed for 100 different nucleic acid configurations sampled from MD simulations to account for the effect of dynamic disorder arising due to the thermal fluctuations.

The reorganization energy, $\lambda_{ik}$, has two parts: inner sphere reorganization energy and outer sphere reorganization energy. Inner sphere reorganization energy considers the change in nuclear degrees of freedom when the charge transfer takes place between one charge hopping site to another. This is defined as:

$$\lambda_{ik}^{int} = U_i^{nC} - U_i^{nN} + U_k^{cN} - U_k^{cC} \qquad (3)$$

$U_i^{nC}$ ($U_i^{cN}$) is the internal energy of neutral (charged) base in charged (neutral) state geometry. $U_i^{nN}$ ($U_i^{cC}$) is the internal energy of neutral (charged) base in neutral (charged) state geometry.

The reorganization of the environment as the charge transfer occurs is considered using the outer sphere reorganization. This has been taken as a parameter (0 eV) in our calculations as the surrounding conditions for both native and oxidized dsDNA are same for all the calculations.

$\Delta G_{ik}^{ext}$ is the contribution due to the external electric field, taken as the potential difference between the two hopping sites in our calculations. We take uniform distribution of potential between the base pairs, i.e. consecutive base pairs will have a potential difference of (V/(N-1)), while bases of same pairs will have zero potential difference i.e.

$\Delta G_{ik}^{ext} = \frac{V}{N-1}$, for consecutive bases along the helical axis of DNA along positive volt.

$\qquad$ 0, for base pairs at the same level along helical axis. $\qquad (4)$

$\qquad -(\frac{V}{N-1})$, for consecutive bases along the helical axis of DNA or along negative volt.

N is the number of base-pairs.

Whereas,

$$\Delta G_{ik}^{int} = U_i^{cC} - U_i^{nN} + U_k^{cC} - U_k^{nN} \tag{5}$$

Where, $U_i^{cC}(U_i^{nN})$ is the internal energy of base in the charged (neutral) state and geometry.

The calculation of transfer integrals and reorganization energies are performed using density functional theory (DFT) which have been carried out with M062X/6-31g(d) functional level of theory using Gaussian09 Software package. The effect of solvation arising due to the surrounding water medium of the base pairs is considered using polarizable continuum model[60] (PCM) in this work. It is noteworthy to mention here that explicit representation of the surrounding environment has also been done using QM/MM methods in previous CT studies of DNA[61,62]. However, since both the native and oxidized dsDNA are immersed in charge neutral TIP3P water box of similar dimensions, their environment is similar[36]. Thus, changing the method of representing external environment should not drastically affect the relative CT properties of native and oxidized dsDNA. Therefore, we use implicit solvation method which has been demonstrated to approximate the DNA environment with sufficient accuracy in previous DNA CT works[45,63,64]. VOTCA-CTP[48] software package is used to calculate the transfer integral values for all possible base pairs.

## 2.3 Kinetic Monte Carlo (KMC)

Once we obtain all the hopping rates, we employ KMC algorithm to simulate the charge dynamics of the system by solving the master equation containing the probabilities of a charge hopping site to hold charge. In this scheme, initially a unit charge is assigned to a random hopping site $i$ and this instant is taken as the initial time, i.e., $t = 0$. The waiting time is then computed using the following relation:

$$\tau = -\omega_i^{-1} \ln(r_1) \tag{6}$$

Here, $\omega_i = \Sigma_{k=1}^{N} \omega_{ik}$ is the sum of all the hopping rates for the sites available for charge present at site $i$ to hop, $n$ is the number of charge hopping sites available for charge at site $i$, $r_1$ represents a uniform random number between 0 and 1 and $k$ is the index for the available sites for hopping. Once the waiting time is calculated, the total time is then updated as $t = t + \tau$. To decide the site to which the charge has to hop to, the j for which $\frac{\Sigma_k \omega_{ik}}{\omega_i} \leq r_2$ is largest and r2, is chosen where r2 is another uniform random number between 0 and 1. The above condition ensures that the site $k$ is chosen with probability $\frac{\omega_{ik}}{\omega}$. After this, the position of the charge is updated and the above process is repeated, providing the probabilities for each site. The current is then found using the following formula

$$I_{bp} = -e \left[ \Sigma_i \left( P_{b_1} \omega_{b_1 i} - P_i \omega_{i b_1} \right) + \Sigma_i \left( P_{b_2} \omega_{b_2 i} - P_i \omega_{i b_2} \right) \right] \tag{7}$$

Here, e is the unit electric charge, $i$ stands for all the possible hopping sites which are in the direction of flow of current, b₁ and b₂ are the base stacks of base pair bp. Hence mean current is average over all base pairs, $I = <I_{bp}>$.

# Section 3. Results

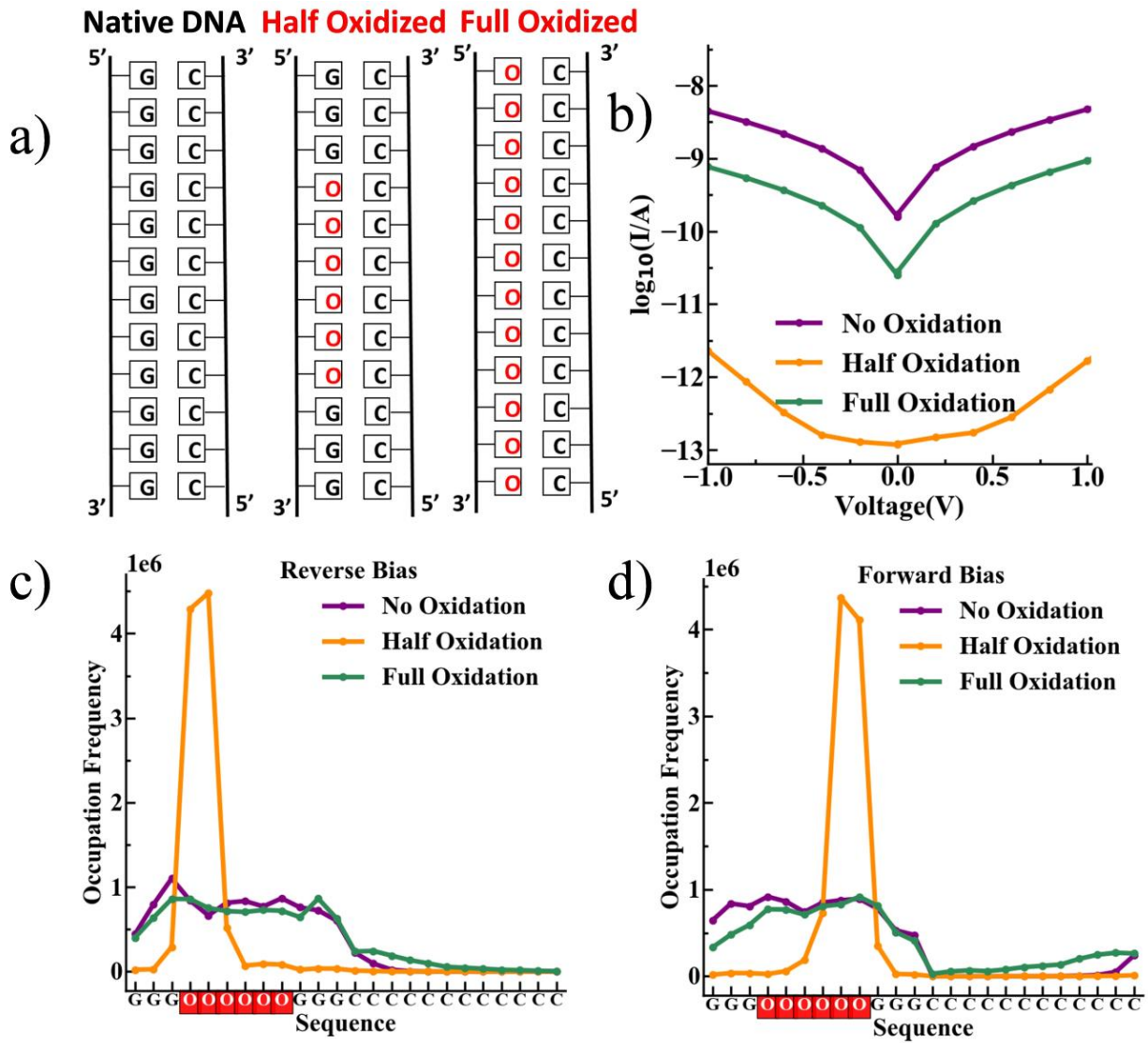

**Fig. 2** a) Schematic representation showing the oxidation of half and all guanines of 12 bp poly-(G) sequence. b) Average I-V characteristics averaged over 100 morphologies of native and oxidized poly-(G) DNA sequence, computed using multiscale methodology. c) The occupation frequency of the hole at different nucleobases of dsDNA under c) reverse bias and d) forward bias. Clearly, the 8oxoG acts as a bottleneck for charge transport in half oxidized dsDNA explaining the I-V characteristics obtained in b).

We have studied the 12 bp poly-G dsDNA sequence by oxidizing all the guanine bases of the DNA chain in one case, while oxidizing only middle half of the sequence in the other case as shown in **Fig. 2(a).** In this dsDNA sequence, we find that the native dsDNA exhibits the highest conductance relative to the oxidized dsDNA, while fully oxidized dsDNA shows higher conductance than half-oxidized dsDNA. This can be explained based on the charge dynamics in different DNA sequences (**Fig. 2(c)-(d)**). As shown in **Fig. 2(c) and 2(d)**, the charge gets trapped in the half-oxidized DNA at the point where the charge must hop between the oxidized and the native guanine base. However, for a native dsDNA or a fully oxidized DNA, such consecutive sequence mismatch is not present. It is precisely the emergence of this charge trapping in 8oxG bases what explains, from a molecular point of view, the lower conductivity in half-oxidized poly-G dsDNA. Indeed, such trapping of hole at the 8oxoG sites has been reported earlier and has found a huge biological significance[65]. Based on experimental evidence, it has been demonstrated that the 8oxoG in dsDNA sequence acts as a hole sink and saves the rest of the genome from further oxidative damage[66]. Our comprehensive analysis based on KMC simulations shows that the hole indeed gets trapped at the 8oxoG sites giving rise to the concomitant effect of lower conductivity in oxidized DNA.

The energetics of 8oxoG and native guanine bases are largely responsible for the trapping of charge in oxidized dsDNA. To verify this, we have computed and compared the reorganization energy and the site energy differences of both guanine and 8oxoG with other nucleobases using M062X level of theory using 6-31g(d) basis set. We find that 8oxoG requires significantly higher reorganization energies and site energy differences to hop to other nucleobases, compared to that required for the native guanine for the same (**Table 1**). This energy difference reduces the probability of hopping of oxidized guanine, and hence the charge gets trapped. To make sure this observation is not an artifact of the basis sets used for the calculation of the energies, we use different basis sets as shown in Table 1 and find that the relative trend remains same for the oxidized and native guanine molecules.

Having built a thorough understanding of the charge dynamics and energetics in native and oxidized dsDNA, we then ask the question whether this knowledge can be used for molecular electronic applications. For this, we study the Drew-Dickerson dsDNA sequence, and replace the guanine nucleobases at different locations and concentrations, as explained in **Fig. 1(e)**. We will now discuss the feasibility of 8oxoG to construct molecular rectifiers using Drew-Dickerson dsDNA as a case study.

**Table 1:** Reorganization Energy and Site Energy Difference of Guanine and 8oxoG to different nucleobases for different basis sets.

| Basis Set: 6-31 g(d) | | | | | | |
|---|---|---|---|---|---|---|
| | | To A | To T | To C | To G | To O |
| **Reorganization Energy (in eV)** | From G | 0.615 | 0.595 | 0.502 | 0.684 | 0.678 |
| | From O | 0.64 | 0.62 | 0.527 | 0.709 | 0.703 |
| **Site Energy Difference (in eV)** | From G | -0.408 | -0.774 | 0.769 | 0.0 | 0.355 |
| | From O | -0.763 | 1.129 | -1.124 | -0.355 | 0.0 |
| **Basis Set: 6-31 g(d,p)** | | | | | | |
| | | To A | To T | To C | To G | To O |
| **Reorganization Energy (in eV)** | From G | 0.57 | 0.59 | 0.49 | 0.66 | 0.66 |
| | From O | 0.60 | 0.62 | 0.52 | 0.687 | 0.686 |
| **Site Energy Difference (in eV)** | From G | -0.41 | -0.78 | 0.78 | 0.0 | 0.35 |
| | From O | -0.77 | 1.13 | -1.13 | -0.35 | 0.0 |
| **Basis Set: 6-311++g(d,p)** | | | | | | |
| | | To A | To T | To C | To G | To O |
| **Reorganization Energy (in eV)** | From G | 0.56 | 0.61 | 0.52 | 0.66 | 0.67 |
| | From O | 0.59 | 0.64 | 0.55 | 0.69 | 0.70 |
| **Site Energy Difference (in eV)** | From G | -0.38 | -0.73 | 0.77 | 0.0 | 0.32 |
| | From O | -0.7 | 1.05 | -1.1 | -0.32 | 0.0 |
| **Basis Set: cc-PVTZ** | | | | | | |

|   |   | To A | To T | To C | To G | To O |
|---|---|------|------|------|------|------|
| **Reorganization Energy (in eV)** | From G | 0.58 | 0.61 | 0.52 | 0.66 | 0.67 |
|   | From O | 0.61 | 0.64 | 0.55 | 0.69 | 0.70 |
| **Site Energy Difference (in eV)** | From G | -0.39 | -0.77 | 0.8 | 0.0 | 0.33 |
|   | From O | -0.72 | 1.11 | -1.13 | -0.33 | 0.0 |

**Basis Set: aug-cc-PVDZ**

|   |   | To A | To T | To C | To G | To O |
|---|---|------|------|------|------|------|
| **Reorganization Energy (in eV)** | From G | 0.54 | 0.58 | 0.50 | 0.63 | 0.64 |
|   | From O | 0.59 | 0.61 | 0.53 | 0.67 | 0.67 |
| **Site Energy Difference (in eV)** | From G | -0.38 | -0.74 | 0.78 | 0.0 | 0.32 |
|   | From O | -0.70 | 1.06 | -1.1 | -0.32 | 0.0 |

In **Fig. 3**, we first study the effect of oxidation of a single guanine nucleobase at different locations in Drew-Dickerson dsDNA. In one case, we replace the $2^{nd}$ nucleobase from the 5' end of dsDNA with 8oxoG (**Fig. 3a**), while in the other case, the nucleobase at $12^{th}$ site is replaced (**Fig. 3e**). **Fig. 3b** and **Fig. 3f** show the comparison of I-V characteristics of native dsDNA with dsDNA oxidized at $2^{nd}$ and $12^{th}$ site, respectively. We find strong rectification behavior in both the cases. While the dsDNA oxidized at $2^{nd}$ site shows 3 orders of magnitude lower conductance ($RR = 1.4 \times 10^{-3}$) in forward bias, a $12^{th}$ site oxidation in dsDNA leads to similar dip in conductance in reverse bias ($RR = 61.2$). It is fascinating to note here that oxidation in only a single guanine base of the whole sequence leads to such drastic changes. The decrease in the conductance certainly comes from the charge dynamics and energetics of nucleobases, which is related to the hopping rates near the 8oxoG bases. In **Fig. 3c** and **Fig. 3d** compare the charge dynamics in the $2^{nd}$ site oxidized dsDNA and native dsDNA for both forward and reverse biases. While the probability of finding the charge at each site is similar in reverse bias (**Fig. 3c**), there is an uptick in the charge occurrence at 8oxoG in forward bias (**Fig. 3d**). This is because in the case of oxidation at $2^{nd}$ site, the oxidation site is closer to positive electrode in forward bias causing a disruption in the flow of holes through the dsDNA. This increased occupation leads to a decrease in conductance in $2^{nd}$ site oxidation in forward bias. Similarly, for $12^{th}$ site oxidation, as the oxidized site is closer to positive electrode in reverse bias, the occupation frequency of charge is higher at 8oxoG in reverse bias (**Fig. 3g**) leading to lesser conductance in reverse bias configuration. In the forward bias, the occupation frequency graph is similar to that of native dsDNA (**Fig. 3h**) leading to a strong overall rectification behavior in the circuit. Thus, by varying the position of oxidation in dsDNA, one can obtain direction-dependent rectification properties in single-molecular circuits made-up of dsDNA molecules.

To further check the effect of concentration of oxidation on dsDNA CT properties, we replace the two guanine bases near the 5' end of one strand with 8oxoG (**Fig. 4a**). Like the single oxidation case, upon inspecting the I-V characteristics, one can immediately observe a very strong rectification behavior in the oxidized dsDNA sequence with a high rectification ratio of around 6 orders of magnitude at 1 V ($RR = 7.8 \: x \: 10^{-6}$). A sharp spike in the occurrence frequency in the forward bias in double oxidized dsDNA marks the trapping of charge on 8oxoG sites. However, the two sequences qualitatively have similar I-V characteristics for the reverse bias. This is because of the asymmetry in the sequence which results in higher energy differences in the forward bias, as the site energy difference increases in the direction of flow of charge. The high rectification ratio of around 3 orders of magnitude makes such dsDNA sequences perfect candidates for molecular switches.

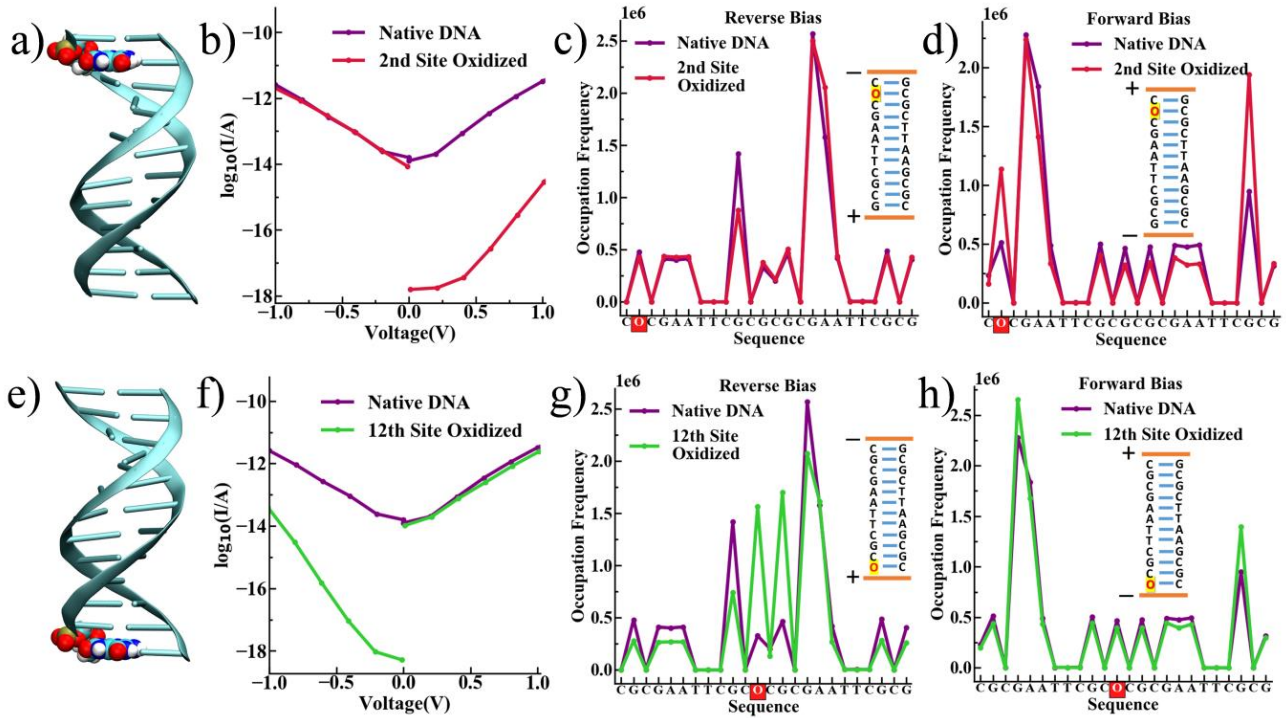

**Fig. 3** Effect of single guanine oxidation on CT properties of dsDNA. a) 3-D structure of Drew-Dickerson dsDNA sequence oxidized at $2^{nd}$ site. The 8oxoG is shown in VDW representation. b) I-V characteristics of the native Drew-Dickerson sequence compared to that oxidized dsDNA with the guanine at $2^{nd}$ site from the end 5' end replaced with 8oxoG. The occupation frequency of the charge during the KMC simulations for c) reverse bias and d) forward bias. Clearly, a sharp increase at the 8oxoG site is seen in the oxidized guanine in the forward bias at 8oxoG, giving rise to rectifying behavior of the molecule. The guanines replaced with 8oxoG are highlighted in red color in the x-axis of the curves. d) 3-D structure of Drew-Dickerson dsDNA sequence oxidized at $12^{th}$ site. f) I-V characteristics of the native Drew-Dickerson sequence and oxidized with the guanine at the $12^{th}$ site from the 5' end of dsDNA replaced with 8oxoG. The occupation frequency of the charge during the KMC simulations for g) reverse bias and h) forward bias.

Next, we replace 4 guanine bases of the two strands at the same end with 8oxoG bases in one case (**Fig. 4(e)**), and opposite ends in the other (**Fig. 4(i)**). In the quadruple oxidation at same end, the I-V characteristics signify that the oxidized dsDNA sequence shows a rectifying behavior with a high rectification ratio of $2.3 \times 10^{-4}$. In the forward bias, the charge gets trapped on the two oxidized Guanine as evident from **Fig. 4(g)** and cannot traverse through these oxidized guanines because of the vast energy difference. In reverse bias, as a voltage is applied at the end of the DNA sequence having the normal guanine bases, the energy disparity between the oxidized guanines in the two strands decreases. Therefore, the I-V characteristics of the native and oxidized DNA are similar in the reverse bias.

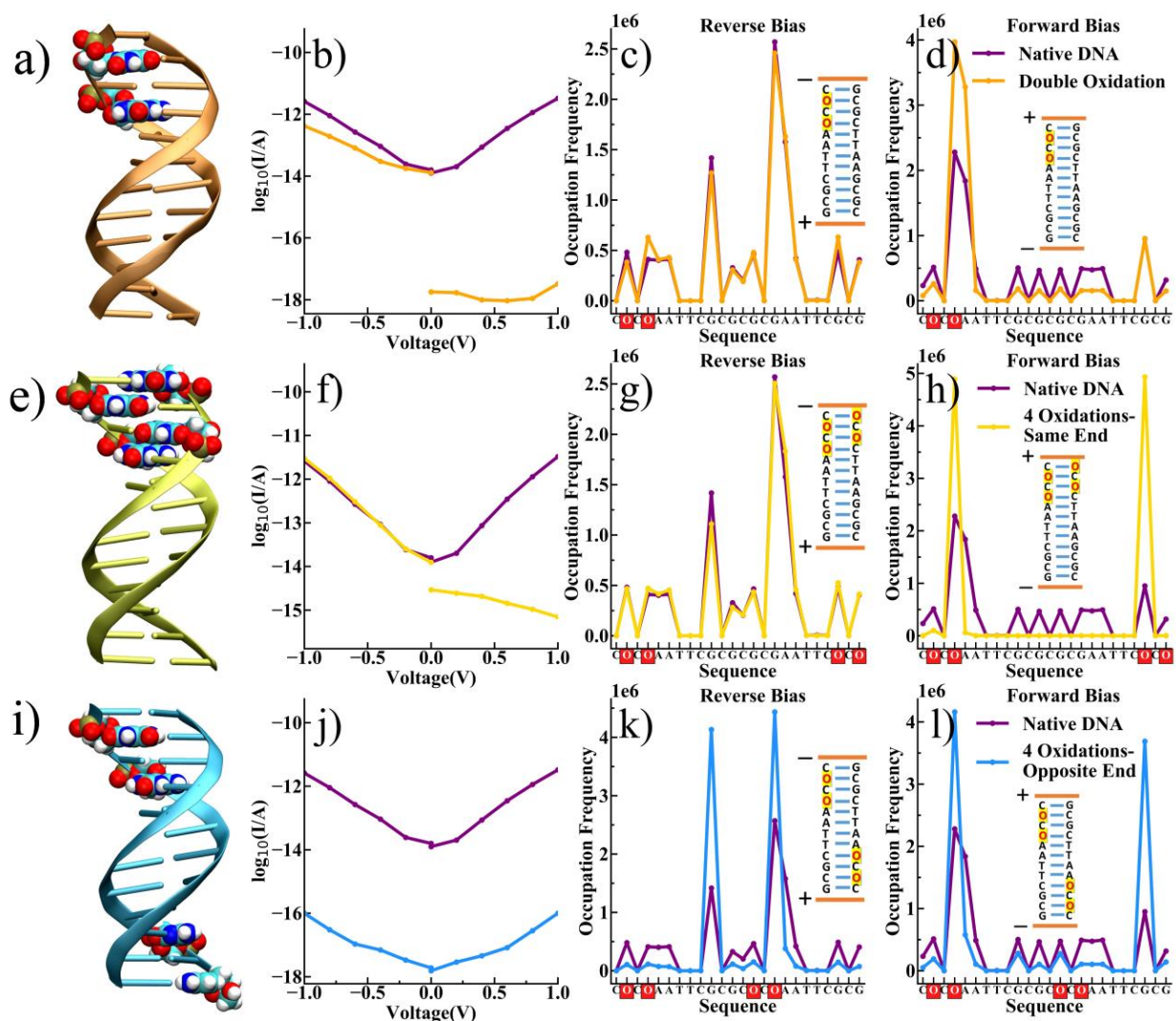

**Fig. 4** Effect of double and quadruple guanine oxidations on CT properties of Drew-Dickerson dsDNA. (a) 3-D structure of double oxidized Drew-Dickerson dsDNA at 2$^{nd}$ and 4$^{th}$ site. The 8oxoG is shown in VDW representation. (b) I-V characteristics of the double oxidized dsDNA. The occupation frequency of the charge during the KMC simulations for (c) reverse bias and (d) forward bias for double oxidation. (e) 3-D structure of quadruple oxidized Drew-Dickerson dsDNA at same end. (f) I-V characteristics and the occupation frequency of the charge during the KMC simulations for (g) reverse bias and (h) forward bias for quadruple oxidation at same end. (i) 3-D structure of quadruple oxidized Drew-Dickerson dsDNA at opposite ends. (f) I-V characteristics and the occupation frequency of the charge during the KMC simulations for (g) reverse bias and (h) forward bias for quadruple oxidation at opposite ends.

In quadruple oxidation at opposite ends, the dsDNA conductance decreases up to 2 orders of magnitude upon oxidation in both forward and reverse bias. This is because of the trapping of charge at the oxidized guanines in both reverse (**Fig. 4 (k)**) and forward (**Fig. 4 (l)**) bias conditions.

However, as both the 5'-3' strands of the dsDNA sequence are symmetric to each other in this case, no rectification behavior is seen.

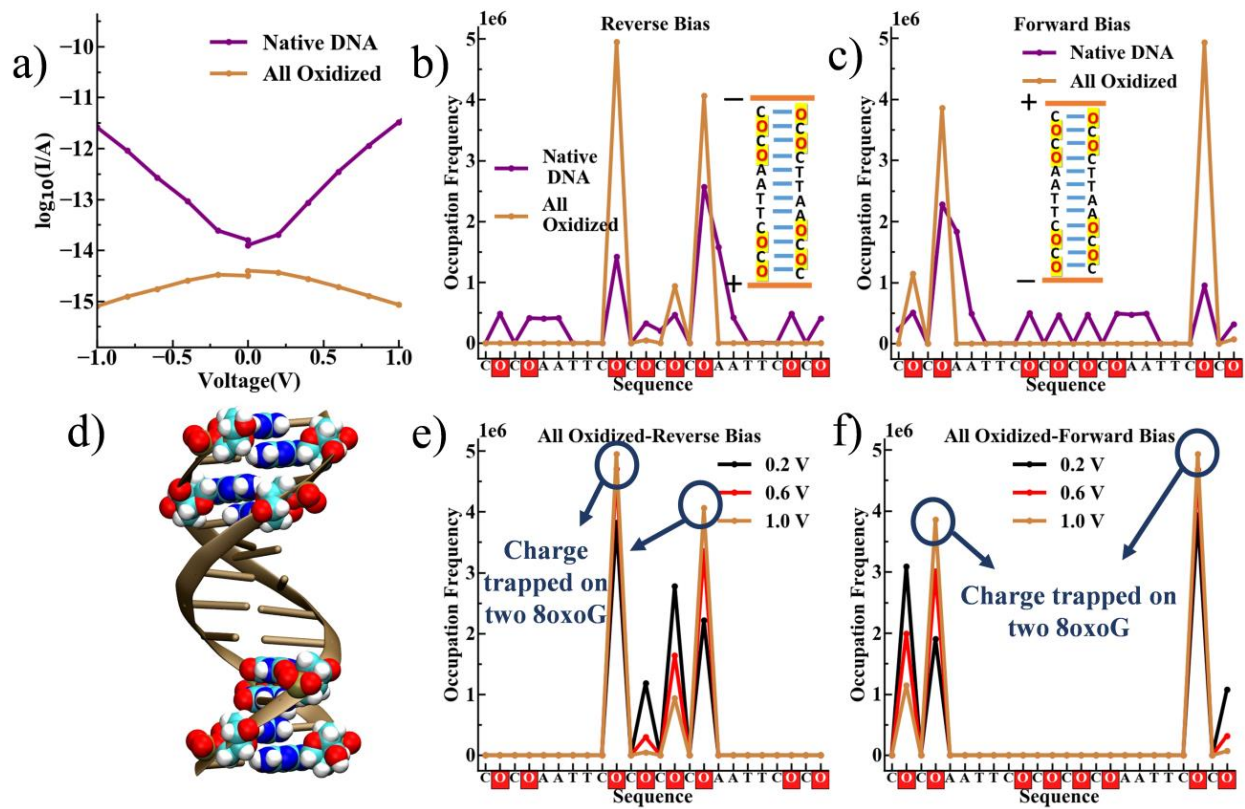

**Fig. 5** Effect of replacing all guanines of Drew-Dickerson sequence with 8oxoG on its CT properties. a) I-V characteristics of the native Drew-Dickerson sequence compared to that oxidized dsDNA with all the guanines at replaced with 8oxoG. The occupation frequency of the charge during the KMC simulations for b) reverse bias and c) forward bias. d) 3-D structure of Drew-Dickerson dsDNA sequence with all the guanines oxidized. The 8oxoGs are shown in VDW representation. The variation of the occupation frequency for different applied voltages in e) reverse bias and f) forward bias. Clearly, as the voltage increases, the charge gets more and more trapped on the two 8oxoGs in all-oxidized dsDNA, leading to negative differential resistance phenomena.

Finally, we study the Drew-Dickerson sequence with all the guanine bases oxidized to 8oxoG (**Fig. 5**). We find that the conductance decreases by an order of magnitude and the trapping of charge at the oxidized guanine sites is higher compared to the native dsDNA (**Fig. 5(a)**). Notably, since the sequence is completely symmetric in this case, the charge gets highly trapped on the 8oxoG sites. It is noteworthy to mention here that the charge gets trapped on the same guanine locations in the sequence, where the occupation frequency was higher in the native dsDNA case as well (**Fig. 5(b)-(c)**).

From **Fig. 5(a)**, one can clearly see that the conductance of fully oxidized Drew-Dickerson sequence shows traits of negative differential resistance (NDR), i.e., the conductance decreases with increasing voltage. To understand the reason behind this observation, we present the occupation frequency graph at different applied potential in reverse bias (Fig. 5e) and forward bias (**Fig. 5f**). We see that as the voltage increases, the charge gets more and more trapped on two 8oxoGs in both the biases. This is because as the potential bias increases, the difference between the site energies of 8oxoG and other nucleobases increases, and the charge has nowhere to escape than getting trapped at two 8oxoGs. Similar traits of NDR can be seen in the forward bias configuration of quadruple oxidation at same end (**Fig. 4f**). Here as well, as seen from **Fig. 4h**, the charge gets trapped at two 8oxoGs, which increases as the potential bias increases. Thus, a complete understanding of the flow of charge in dsDNA molecules can be made by analyzing the charge dynamics. This provides a unique tool to control/tune the flow of charge in a dsDNA sequence.

## Conclusions

In conclusion, we have presented a novel strategy to construct a molecular rectifier with record rectification ratio of $10^6$ using oxidatively damaged dsDNA molecules. The ability to harness the power of rectifying ability of oxidized dsDNA opens doors to numerous practical applications such as molecular switches, bioelectronic sensors, molecular diodes, and so on. Using a multiscale methodology, that combines MD, DFT and KMC simulations, we find that an oxidized guanine acts as a charge trapping barrier leading to lesser electronic conduction in oxidized dsDNA relative to the corresponding native dsDNA. This work provides a fundamental understanding of the charge dynamics in oxidatively damaged dsDNA sequences and a novel strategy to use DNA molecules in electronic applications. This study will motivate further experimental works to explore the tunable rectifying properties of DNA using state-of-the-art single molecule experimental techniques[67] such as scanning tunneling microscopy-break junction (STM-BJ) and mechanically controllable-break junction (MCBJ) setups. Having a huge biological significance, an elementary understanding of the CT processes in oxidized dsDNA is essential to study their role in countless cell functions and will ultimately aid to cure numerous diseases.

# Conflicts of interest

There are no conflicts of interest to declare.

# Acknowledgments

A.A. and S.N. thank MHRD, India and CSIR, India respectively for the research fellowship. Authors also acknowledge DST, India, for the computational support through TUE-CMS machine, IISc.